\documentclass[runningheads]{llncs}

\usepackage[notheorems]{dgleich-math}

\usepackage{graphicx,epsfig}
\usepackage{graphics,dcolumn,bm,epic,eepic,float}
\usepackage{amssymb,amsmath,multirow,rotate,color}
\usepackage{subfigure}
\usepackage{epstopdf}
\usepackage{tabularx}
\usepackage{booktabs}

\usepackage{paralist}

\usepackage{url}
\usepackage{amssymb}
\usepackage{graphicx}
\usepackage{helvet}
\usepackage{courier}
\usepackage{subfigure}
\usepackage{amsmath}
\usepackage{multirow}
\usepackage{comment}
\usepackage{amssymb}
\usepackage{multirow}
\usepackage{color}
\usepackage{algorithm}
\usepackage{algorithmic}
\usepackage{graphics}
\usepackage{rotating}
\usepackage{wrapfig}
\usepackage{hyperref}
\usepackage{xspace}
\hypersetup{colorlinks=true,urlcolor=blue,citecolor=blue,linkcolor=blue}

\usepackage{amsmath}
\usepackage{amssymb}
\usepackage{textcase}
\usepackage{soul}
\usepackage{dashrule}

\usepackage{url}

\newcommand{\be}{\begin{equation}}
\newcommand{\ee}{\end{equation}}
\newcommand{\bea}{\begin{eqnarray}}
\newcommand{\eea}{\end{eqnarray}}
\newcommand{\bit}{\begin{itemize}}
\newcommand{\eit}{\end{itemize}}

\newcommand{\cm}[1]{}
\newcommand{\jt}[1]{}

\usepackage{microtype}

\begin{document}

\mainmatter

\title{Dynamic PageRank using \\Evolving Teleportation}

\titlerunning{Dynamic PageRank}

\author{Ryan A.~Rossi
\and David F.~Gleich
}

\authorrunning{R.~A.~Rossi and D.~F.~Gleich}

\institute{Purdue University\\
Department of Computer Science\\
305 N. University St., West Lafayette, IN 47906\\
\{rrossi, dgleich\}@purdue.edu
}

\toctitle{Lecture Notes in Computer Science}
\tocauthor{Authors' Instructions}
\maketitle

\begin{abstract}
The importance of nodes in a network constantly fluctuates based on changes in the network structure as well as changes in external interest.
We propose an evolving teleportation adaptation of the PageRank method to capture how changes in external interest influence the importance of a node.  This framework seamlessly generalizes PageRank because the importance of a node will converge to the PageRank values if the external influence stops changing.  We demonstrate the effectiveness of the evolving teleportation on the Wikipedia graph and the Twitter social network.  The external interest is given by the number of hourly visitors to each page and the number of monthly tweets for each user.
\end{abstract}

\section{Introduction}
Finding important nodes in a graph is a key task in a variety of applications:  search engines~\cite{page1998pagerank,kleinberg1999authoritative}, network science~\cite{katz1953new,bonacich1987power,freeman1979centrality}, and bioinformatics~\cite{suzuki2001identification,morrison2005-generank}, among many others.
By and large, these are global measures of node importance and 
one of the most well-studied measures is PageRank~\cite{page1998pagerank,langville2006-book}.

PageRank computes the importance of each node in a directed graph under a random surfer model.
When at a node, the random surfer can either:
\begin{compactenum}
\item transition to a new node from the set of out-edges, or
\item do something else (e.g., execute a search query, use a bookmark).
\end{compactenum}
The probability that the surfer performs the first action is known as the damping parameter in PageRank. We use $\alpha$ to denote the damping parameter.  The second action is called teleporting and is modeled by the surfer picking a node at random according to a distribution called the teleportation distribution vector or personalization vector. These choices only depend on the current node and, consequently, define a Markov chain.  This PageRank Markov chain always has a unique stationary distribution for any $0 \le \alpha < 1$. The importance of a node is proportional to its stationary distribution in this Markov chain.  Thus, the computation is governed by the graph, a teleportation parameter $\alpha$, and a teleportation distribution vector.

The PageRank score is a simple model for the importance of a node in a graph, and there are many variations that may yield more useful scores (for instance~\cite{mathieu2004-backbutton} models a random walk with a back button).  A common complaint about PageRank models is that they are only defined for static graphs.  Motivated by the idea of studying PageRank with dynamic graphs, we formulate a dynamic PageRank model for a static graph with a time-dependent, or evolving, teleportation vector.  
Intuitively, the teleportation distribution changes based on human dynamics such as recent news and seasonal preferences. 
For example, in our forthcoming experiments (Section~\ref{sec:results}), the time-dependent vector is the number of hourly page visits for each page from Wikipedia.  We derive the model and algorithms for this dynamic version of PageRank in Section~\ref{sec:dynamic-pagerank}.  The resulting algorithms scale to large graphs.  Moreover, we show that the new model is a generalization of PageRank in the sense that \emph{if the time-dependent vector stops changing} then \emph{our dynamic score vector converges to the standard PageRank score}. 

We make our code and data available in the spirit of reproducible research:

\centerline{ \footnotesize \url{http://www.cs.purdue.edu/homes/dgleich/codes/dynsyspr-waw}}

\section{PageRank notation}\label{sec:notation}
In order to place our work in context, we first introduce some
notation.   Let $\mA$ be the adjacency matrix for a graph where
$A_{i,j}$ denotes an edge from node $i$ to node $j$.  In
order to avoid a proliferation of transposes, we define $\mP$
as the transposed transition matrix for a random-walk on
a graph: 
\[ P_{j,i} = \text{ probability of transitioning from node $i$ to node $j$. } \]
Hence, the matrix $\mP$ is \emph{column-stochastic} instead of
row-stochastic, which is the standard in probability theory.
Throughout this manuscript, we utilize uniform random-walks on a graph, 
in which case $\mP = \mA^T \mD^{-1}$ where $\mD$ is a diagonal matrix with
the degree of each node on the diagonal.  However, none of the theory
is restricted to this type of random walk and any 
column-stochastic matrix will do.  The PageRank vector $\vx$
is the solution of the linear system: 
\[ ( \mI - \alpha \mP) \vx = (1-\alpha) \vv \]
for any $0 \le \alpha < 1$ and any teleportation distribution 
vector $\vv$ such that $v_i \ge 0$ and $\sum v_i = 1$.
Table~\ref{table:notation} summarizes these notation conventions,
and has a few other elements that will be discussed in the
forthcoming sections.

\begin{table}[t!]
\caption{Summary of notation. Matrices are bold, upright roman letters; vectors are bold, lowercase roman letters; and scalars are unbolded roman or greek letters.}
\label{table:notation}
\centering 
\small
\begin{tabularx}{\linewidth}{rX} 
\toprule
$n$ & number of nodes in a graph  \\
$\ve$ & the vector of all ones \\
$\mP$ & column stochastic matrix  \\
$\alpha$ & damping parameter in PageRank  \\
$\vv$ & teleportation distribution vector \\
$\vx$ & solution to the PageRank computation  \\
\midrule
$\vv(t)$ & a teleportation distribution vector at time $t$  \\
$\vx(t)$ & solution to the Dynamic PageRank computation for time $t$ \\
$\theta$ & decay parameter for time-series smoothing \\
\bottomrule
\end{tabularx}
\end{table}

\section{Dynamic and Evolving Rankings}\label{sec:related-work}
The PageRank literature is vast, and we now survey some 
of the other ideas related to incorporating graph dynamics
into a PageRank vector, more general models for studying
dynamic graphs, and updating PageRank vectors.

Our proposed method is related to
changing the teleportation vector in the power method
as its being computed.  Bianchini et 
al.~\cite{bianchini2005-inside-pagerank} noted
that the power method would still converge if
either the graph or the vector $\vv$ changed during
the method, albeit to a new solution given
by the new vector or graph.  Our method capitalizes
on a closely related idea and we utilize the 
intermediate quantities explicitly.  Another related
idea is the Online Page Importance Computation 
(OPIC)~\cite{abiteboul2003adaptive}, which integrates
a PageRank-like computation \emph{with} a crawling
process.  The method does nothing special if
a node has changed when it is crawled again.  A more
detailed study of how PageRank values evolve during
a web-crawl was done by Boldi et 
al.~\cite{boldi2005-incremental-pagerank}.  
Other work has approximated PageRank on graph 
streams~\cite{das2008estimating}.

Outside
of the context of web-ranking, 
O'Madadhain and Smyth propose EventRank~\cite{o2005eventrank},
a method of ranking nodes in dynamic graphs, that
uses the PageRank propagation equations for a sequence of graphs.
We utilize the same idea but place it within the context
of a dynamical system.  

While we described PageRank in terms of a random-surfer
model above, another characterization of PageRank is that
it is a sum of damped transitions: 
\[ \vx = (1-\alpha) \sum_{k=0}^\infty (\alpha \mP)^k \vv. \]
These transitions are a type of probabilistic walk and 
Grindrod et al.~\cite{grindrod2011communicability}
introduced the related notion of dynamic walks for
dynamic graphs.

In the context of popularity dynamics~\cite{ratkiewicz2010characterizing}, our method captures how changes in external interest influence the popularity of nodes and the nodes linked to these nodes in an implicit fashion.
Our work is also related to modeling human dynamics, namely, how humans change their behavior when exposed to rapidly changing or unfamiliar conditions~\cite{bagrow2011collective}. 
In one instance, our method shows the important topics and ideas relevant to humans before and after one of the largest Australian Earthquakes.

In closing, we wish to note that our proposed method \emph{does not involve}
updating the PageRank vector, a related problem which has
received considerable attention~\cite{chien2004link,langville04-iad}.
Nor is it related to tensor methods for dynamic graph
data~\cite{Sun-2006-tensor,Dunlavy-2011-temporal}.

\section{PageRank with Dynamic Teleportation}\label{sec:dynamic-pagerank}
In order to incorporate dynamics into PageRank, we reformulate a standard 
PageRank algorithm in terms of changes to the PageRank values for each page.
This step allows us to state PageRank as a dynamical system, 
in which case we can easily incorporate
changes into the vector.

The standard PageRank algorithm is the classical Richardson iteration: 
\[ \vx\itn{k+1} = \alpha \mP \vx \itn{k} + (1-\alpha) \vv. \]
(Note that this iteration is identical to the power method for the PageRank
Markov chain.)
By rearranging this equation into a difference form, we have 
\[ \Delta \vx\itn{k} = \vx\itn{k+1} - \vx\itn{k} = \alpha \mP \vx\itn{k} + (1-\alpha) \vv - \vx\itn{k} = (1-\alpha) \vv - (\eye - \alpha \mP) \vx\itn{k}. \]
Thus, changes in the PageRank values at a node \emph{evolve} based on the value $(1-\alpha) \vv - (\eye - \alpha \mP) \vx\itn{k}$.  We reinterpret this update as a continuous time dynamical system:
\begin{equation} \label{eq:pr-dynamical}
 \vx'(t) = (1-\alpha) \vv - (\eye - \alpha \mP) \vx(t). 
\end{equation}
Other iterative methods also give rise to related dynamical systems, as utilized by~\cite{Embree-2009-dynamical} for studying eigenvalue solvers.  

In the dynamic teleportation model, $\vv$ is no longer fixed, but is instead a function of time $\vv(t)$:
\begin{equation} \label{eq:dynamic-tele-pr} 
 \vx'(t) = (1-\alpha) \vv(t) - (\eye - \alpha \mP) \vx(t). 
\end{equation}
Note that this means the PageRank values $\vx(t)$ may not ``settle'' or converge.  We see this as a feature of the new model as we plan to utilize information from the evolution and changes in the PageRank values.  

Standard texts on dynamical system show that the solution $\vx(t)$ is:
\[ \vx(t) = \exp[ -(\eye - \alpha \mP)t ] \vx(0) + (1-\alpha) \int_0^t \exp[-(\eye - \alpha \mP) (t-\tau)] \vv(\tau) \, d \tau. \]
If $\vv(t) = \vv$ is constant with respect to time, then
\[ \int_0^t \exp[-(\eye - \alpha \mP) (t-\tau)] \vv(\tau) \, d \tau = (\eye - \alpha \mP)^{-1} \vv - \exp[ -(\eye - \alpha \mP)t ] (\eye - \alpha \mP)^{-1} \vv. \]
Hence, for constant $\vv(t)$:
\[
\vx(t) = \exp[ -(\eye - \alpha \mP)t ] (\vx(0) - \vx) + \vx,
\]
where $\vx$ is the solution to static PageRank: $(\eye - \alpha \mP) \vx = (1-\alpha)\vv$.
Because all the eigenvalues of $-(\eye - \alpha \mP) < 0$, the matrix exponential terms disappear in a sufficiently long time horizon.
Thus, when $\vv(t) = \vv$, nothing has changed. We recover the original PageRank vector $\vx$ as the steady-state solution:
\[ \lim_{t \to \infty} \vx(t) = \vx \text{ the PageRank vector. } \]
This derivation shows that dynamic teleportation PageRank is a generalization of the PageRank vector.

\subsection{Algorithms}
In order to compute the time-sequence of PageRank values $\vx(t)$, we can evolve the dynamical system \eqref{eq:pr-dynamical} using any standard method, for instance a forward Euler or a Runge-Kutta method.  At the moment, we only use the forward Euler method for simplicity.  This method lacks high accuracy, but is fast and straightforward.  Forward Euler approximates the derivative with a first order Taylor approximation:
\[ \vx'(t) \approx \frac{\vx(t+h) - \vx(t)}{h}, \]
and then uses that approximation to estimate the value at a short time-step in the future: 
\[ \vx(t+h) = \vx(t) + h \left[ (1-\alpha) \vv(t) - (\eye - \alpha \mP) \vx(t) \right] . \]
Note that if $h = 1$ and $\vv(t) = \vv$ for all $t$, then this update becomes the original Richardson iteration.
A summary of this derivation as a formal algorithm to compute a dynamic teleportation PageRank time series is given by Figure~\ref{fig:alg-dynpr}.

\begin{figure}[t]
\centering
\begin{algorithmic}
 \REQUIRE ~\newline
  a graph $G=(V,E)$ and a procedure to compute $\mP \vx$ for this graph \newline
  a maximum time $t_{\max}$ \newline
  a function to return $\vv(t)$ for any $0 \le t \le t_{\max}$ \newline
  a damping parameter $\alpha$ \newline
  a time-step $h$
 \ENSURE $\mX$ where the $k$th column of $\mX$ is $\vx(0 + kh)$ 
  for all $1 \le k \le t_{\max}/h$ (or any
  desired subset of these values) \newline
 \STATE $t \leftarrow 0$; $k = 1$
 \STATE $\vx(0) \leftarrow \vv(0)$ (or any other desired initial condition)
 \WHILE {$t \le t_{\max}-h$}
   \STATE $\vx(t+h) \leftarrow \vx(t) + h \left[ (1-\alpha) \vv(t) - (\eye - \alpha \mP) \vx(t) \right] $
   \STATE $\mX(:,k) \leftarrow \vx(t+h)$
   \STATE $t \leftarrow t + h$; $k \leftarrow k + 1$
 \ENDWHILE  
\end{algorithmic}
\caption{In order to compute a sequence of dynamic teleportation PageRank values, we utilize a forward Euler method for the dynamical system: $\vx'(t) = (1-\alpha) \vv(t) - (\eye - \alpha \mP) \vx(t)$. The resulting procedure looks remarkably similar to the standard Richardson iteration to compute a PageRank vector.  A key difference is that there is no notion of convergence.}\label{fig:alg-dynpr}
\end{figure}

\subsection{Discussion of the algorithm \textit{\&} practical issues}

First, the algorithm we propose easily scales to large networks.  This isn't surprising given its close relationship to the Richardson method for PageRank.  The major expense is the set of $t_{\max}/h$ matrix-vector products with $\mP$ -- all of the other work is linear in the number of nodes.  It could also be used in a distributed setting if any distributed matrix-vector product is available. 

In one sense, the forward Euler method is simply running a power method, but changing the vector $\vv$ at every iteration.  However, we derived this method based on evolving~\eqref{eq:dynamic-tele-pr}.  Thus, by studying the relationship between~\eqref{eq:dynamic-tele-pr} and the algorithm in Figure~\ref{fig:alg-dynpr}, we can understand the underlying problem solved by  changing the teleportation vector while running the power method.  Consequently, we gain additional flexibility in adapting~\eqref{eq:dynamic-tele-pr} to problems.

Thus far, we also have not discussed how to set $\vv(t)$ beyond the brief allusion at the beginning that the dynamic teleportation will be based on Wikipedia pageviews.  When we apply the dynamic teleportation PageRank model, we need to pick a relationship between the time-scale of the dynamical system~\eqref{eq:dynamic-tele-pr} and the time-scale in the underlying application.  For instance, does $\vx(1)$ correspond to the PageRank values after a second, an hour, a day?  There is no ``correct'' answer and the relationship has implications on the final model.  

Suppose that we set $\alpha = 0.85$, $h=1$, and that $t=1$ is a minute of time in the application.  If we have hourly data on Wikipedia pageviews, then the above algorithm will compute $60$ iterations of the power-method between each hour.  If we further use the incredibly simple model that $\vv(t)$ changes each hour as we get new data, then the forward Euler method is essentially equivalent to running the power-method to convergence after $\vv$ changes on the hour.  (They are essentially equivalent in the sense that PageRank will have converged to a 1-norm error of $10^{-4}$ in about 60 iterations.)  If, instead, we set $\alpha=0.85$, $h=1$, and $t=1$ to be 20 minutes of time in the application, then we will do 3 iterations of the power method after each hourly change.

In the preceding discussion of the algorithm, we hypothesized that $\vv(t)$ changes at fixed intervals based on incoming data.  A better idea is to smooth out these ``jumps'' using an exponentially weighted moving average. We plan to investigate this in the future.

\subsection{Ranking from Time-Series} \label{sec:diff-ranking}
The above equations provide a time-series of dynamic PageRank vectors for the nodes, denoted formally as $\vx(t), 0 \le t \le t_{\max}$.  Most applications, however, want a single score, or small set of scores, to characterize the importance of a node.  We now discuss a few ways in which these time series give rise to scores. 
Reference~\cite{o2005eventrank} used similar ideas to extract a single score from a time-series.

\paragraph{Transient Rank.} We call the instantaneous values of $\vx(t)$ a node's \emph{transient} rank.  This score gives the importance of a node at a particular time.

\paragraph{Summary \textit{\&} Cumulative Rank.}  Any summary function $s$ of the time series, such as the integral, average, minimum, maximum, variance, is a single score that encompasses the entire interval $[0,t_{\max}]$.  We utilize the \emph{cumulative rank} in the forthcoming experiments: 
\[ \vc = \int_0^{t_{\max}} \vx(t) \, dt \approx h \mX \ve. \]

\paragraph{Difference Rank.} A node's difference rank is the difference between its maximum and minimum rank over all time: 
\[ \vd = \max_t[\vx(t)] - \min_t[\vx(t)]. \]
Nodes with high difference rank should reflect important events that occurred within the range $[0, t_{\max}]$.
The underlying intuition is that normal nodes are the pages where the Dynamic PageRanks do not change much. While the pages that have large differences in their time-series of PageRanks are topics or news that went viral or becomes popular over time. See Section~\ref{sec:results} for more details and Figure~\ref{fig:evolving-dpr} for examples such as Rihanna, PricewaterhouseCoopers, Watchmen, and American Idol (season 8).

Having a variety of different scores derived from the same data frequently helps when using these scores as features in a prediction or learning task~\cite{becchetti2008-spam,constantine2009random}.

\subsection{Clustering the Time-Series}

After applying our forward Euler based algorithm, we have sampled an approximation of this time-series: $\mX = \{\vx(hk) : k = 1, \ldots, t_{\max}/h \}$. By \emph{clustering} these discrete time-series, we can automatically discover patterns such as increasing or decreasing trends, periodic bursts at certain times of the year, and their ilk.  Our initial experiments were promising but were omitted due to space.

\section{Datasets}\label{sec:datasets}
In both of the following experiments, we set $h=1$, and $t=5$ to represent one period of data -- one hour for Wikipedia and one month for Twitter -- so that we do 5 iterations of the forward Euler method before incorporating the new data.  In each period $\vv(t)$ is normalized to sum to 1, but is otherwise unchanged.

\paragraph{Wikipedia Article Graph and Hourly Pageviews.} Wikipedia provides access to copies of its database~\cite{wikipedia2009}. We downloaded a copy of its database on March 6th, 2009 and extracted an article-by-article link graph, where an article is a page in the main Wikipedia namespace, a category page, or a portal page. All other pages and links were removed. See~\cite{gleich2007three} for more information.

Wikipedia also provides hourly pageviews for each page~\cite{pageviews2009}. These are the number of times a page was viewed for a given hour. These are not unique visits. We downloaded the raw page counts and matched the corresponding page counts to the pages in the Wikipedia graph. We used the page counts starting from March 6, 2009 and moving forward in time.

As an aside, let us note that vertex degrees and cumulated pageviews are uncorrelated with a correlation coefficient of 0.02, indicating that using pageviews will not reinforce any degree bias in the dynamic ranks. In fact, pages with a large number of pageviews may not have high in-degree at all, which provides evidence that pages with large in-degree are not always visited more frequently.

\paragraph{Twitter Social Network and Monthly Tweet Rates.}  We use a follower graph generated by starting with a few seed users and crawling follows links from 2008. We extract the user tweets over time from $2008 - 2009$. A tweet is represented as a tuple $\langle$user, time, tweet$\rangle$. Using the set of tweets, we construct a sequence of vectors to represents the number of tweets for a given month.

\begin{table}[h!]
\caption{Dataset Properties. The pageviews or tweets is denoted as $\vp$.}
\label{table:dataset}
\centering\small
\begin{tabularx}{\linewidth}{ l XX r@{\;\;} X XX } 
\toprule
Dataset &  Nodes & Edges & $t_{\max}$ & Period & Average $p_i$ & Max $p_i$  \\
\midrule
\textsc{wikipedia}  & 4,143,840   &  72,718,664  &  20  &  hours  &  1.3225  &  334,650\\
\textsc{twitter} & 465,022  & 835,424 & 6 &  months &  0.5569  & 1056  \\
\bottomrule
\end{tabularx}
\end{table}

\section{Empirical Results}\label{sec:results}
In this section, we demonstrate the effectiveness of Dynamic PageRank as a method for automatically adapting page importance based on graph structure and external influence by showing that it provides different insights (\S\ref{sec:rank-ts}), finds interesting pages (\S\ref{sec:patterns}), and helps predict pageviews (\S\ref{sec:prediction}).

\subsection{Ranking from Time-Series}\label{sec:rank-ts}
We first use the intersection similarity measure to evaluate the rankings~\cite{Boldi05totalrank}. 
Given two vectors $\vx$ and $\vy$, the intersection similarity metric at $k$ is the average symmetric difference over the top-$j$ sets for each $j \leq k$. If $\mathcal{X}_k$ and $\mathcal{Y}_k$ are the top-$k$ sets for $\vx$ and $\vy$, then
$ \mathrm{isim}_k(\vx,\vy) = \frac{1}{k} \sum_{j=1}^{k} \frac{|\mathcal{X}_j \Delta \mathcal{Y}_j|}{2j} $, 
\noindent 
where $\Delta$ is the symmetric set-difference operation. Identical vectors have an intersection similarity of 0.

\begin{figure}[t!]
\centering
    \includegraphics[width=3.4in]{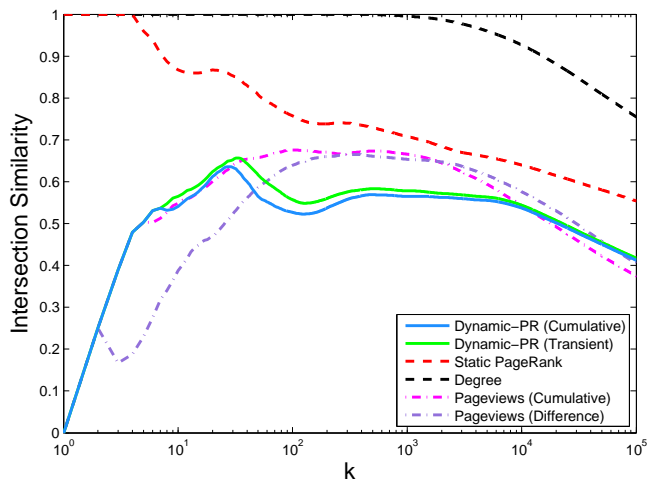}  
   \caption{
Intersection similarity between Dynamic PageRank's difference ranking $\vd$ and the other ranking vectors.
To more appropriately see the differences, we zoom in on the top $10^5$ nodes.
See the discussion in the text.
}
  \label{fig:isim}
\end{figure}

For the Wikipedia graph, Figure~\ref{fig:isim} shows the similarity profile comparing ${\vd}$ (from \S\ref{sec:diff-ranking}) to static PageRank, degree, cumulative pageviews $\vp_c$, maximum pageviews difference $\vp_d$, and two other Dynamic PageRank vectors: transient $\vx(t_{\max})$ and cumulative $\vc$.
The figure suggests that Dynamic PageRank is different from the other measures, even for small values of $k$.  In particular, combining the external influence with the graph appears to produce something new.

\subsection{Top Dynamic Ranks} \label{sec:patterns}
\begin{figure*}[t!]
\includegraphics[width=4.95in]{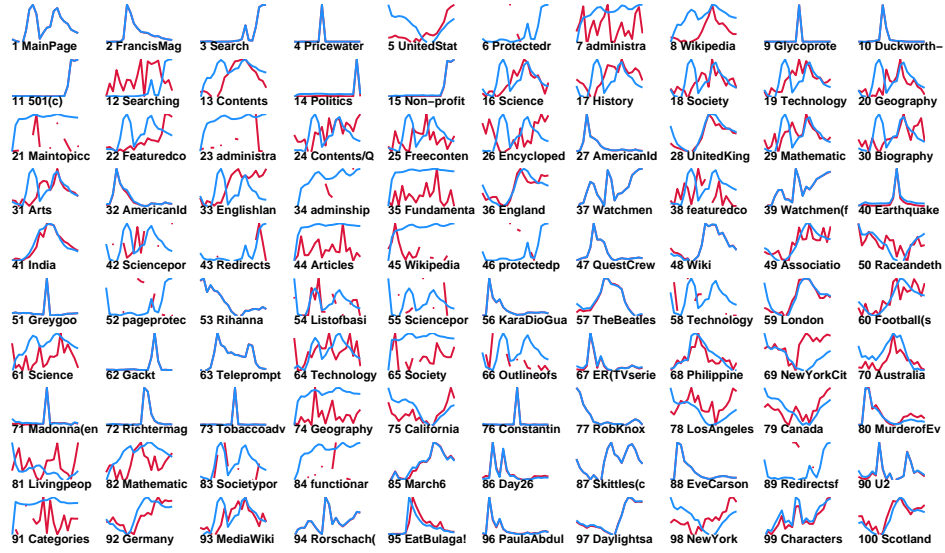}
   \caption{The top-100 Wikipedia pages that fluctuate the most as determined by the difference ranking from our Dynamic PageRank approach. 
    The x-axis represents time (in hours) while the y-axis represents the Dynamic PageRank value. The blue line represents Dynamic PageRank and the red line represents the hourly pageviews.
There exist many interesting time-series patterns such as spikes (40), cyclic/seasonality trends (16-20), and increasing/decreasing trends (39 and 77), among many others. Further analysis and anecdotal evidence was removed due to space.}
  \label{fig:evolving-dpr}
\end{figure*}

Figure~\ref{fig:evolving-dpr} shows the time-series of the top 100 pages by the difference measure. Many of these pages reveal the ability of Dynamic PageRank to mesh the network structure with changes in external interest. This became immediately clear after reviewing significant events from this time period.  
We find pages related to an Australian earthquake (40, 72, 70), a just released movie ``Watchmen'' (94, 39, 99), a famous musician that died (2, 95, 68), recent ``American Idol'' gossip (32, 96, 56), a remembrance of Eve Carson from a contestant on ``American Idol'' (80, 88, 27), news about the murder of a Harry Potter actor (77), and the Skittles social media mishap (87).  These results demonstrate the effectiveness of the Dynamic PageRank to identify interesting pages that pertain to external interest.  The influence of the graph results in the promotion of pages such as Richter magnitude (72).  That page was not in the top 200 from pageviews.

In another study, omitted due to space, we performed a clustering of these time-series to identify pages with similar trends.  
For instance, pages such as Watchmen (37) and Rorschach (94) share strikingly similar patterns. 
These patterns indicate the page that became important first and the amount of traffic or popularity that diffused over time. 

\subsection{Predicting Future Pageviews \textit{\&} Tweets}\label{sec:prediction}
We conclude by studying how well the dynamic PageRank values \emph{predict} future pageviews.
Formally, given a lagged time-series~\cite{ahmed2010empirical}, the goal is to predict the future value $\vp_{t+1}$ (actual pageviews or number of tweets). 
This type of temporal prediction task has many applications, such as actively adapting caches in large database systems, or dynamically recommending pages.

We performed one-step ahead predictions ($t+1$) using linear regression. 
That is, we learn a model of the form:
\[ \left[ \begin{array}{cccccccc}
\bar{\vf}(t-1;\theta) & \bar{\vf}(t-2;\theta) & \ldots  & \bar{\vf}(t-w;\theta)  \end{array} \right]
\; \vb \; \approx \; 
 \begin{array}{c}
\vp(t)  \end{array}\]
where $w$ is the window-size,  and $\bar{\vf}(\cdot;\theta)$ is an exponentially damped moving average computed from either pageviews, dynamic PageRanks, or both.  Using this average is a standard forecasting technique.  
Specifically, the exponentially damped moving average of a time-series feature $\vf(t)$ is: 
\[ \bar{\vf}(t;\theta) = \underbrace{\theta \vf(t)}_{\text{new data}} + \underbrace{(1-\theta)\bar{\vf}(t-1;\theta)}_{\text{old data}}. \]
The exponential factor was $\theta = 0.3$ for Twitter and $\theta = 0.7$ for Wikipedia.  
Due to the scarcity of the data, we used $0.3$ for Twitter since this choice weights past observations more heavily.  In the future, we plan to use cross-validation.
After fitting, the model predicts $\vp(t+1)$ as 
$
\left[ \begin{array}{cccccccc}
\bar{\vf}(t;\theta) & \bar{\vf}(t-1;\theta) & \cdots  & \bar{\vf}(t-w+1;\theta)  \end{array} \right]
\; \vb
$.
To measure the error, we use symmetric Mean Absolute Percentage Error (or sMAPE)~\cite{ahmed2010empirical}.

We study two models.
  
\paragraph{Base Model.} This model uses only the time-series of pageviews  or tweet-rates to predict the future pageviews or number of tweets.

\paragraph{Dynamic PageRank Model.} 
This model uses both the Dynamic PageRank time-series and pageviews to predict the future pageviews.

We evaluate these models for prediction on \textit{stationary} and \textit{non-stationary} time-series. 
Informally, a time-series is weakly stationary if it has properties (mean and covariance) similar to that of the time-shifted time-series. 
We consider the top and bottom 1000 nodes from the difference ranking as nodes that are approximately non-stationary (volatile) and stationary (stable), respectively.  
Table~\ref{table:pv-preds} compares the predictions of the models across time for non-stationary and stationary prediction tasks. 
Our findings indicate that the Dynamic PageRank time-series provides valuable information for forecasting future pageviews.

\begin{table}
\caption{Average SMAPE over all nodes for the two models (lower is better).
We also measure the performance of the models for predicting highly volatile nodes (non-stationary) and nodes with relatively stable behavior (stationary). 
In all cases, the Dynamic PageRank model is more accurate than the base model.}
\label{table:pv-preds}
\centering\small
\begin{tabularx}{\linewidth}{ lXXX}
\toprule
 \textbf{{Dataset}} & \textbf{{Forecasting}} & \textbf{Dynamic PageRank}& \textbf{Base Model}\\
\midrule
\textsc{wikipedia} & 
\textit{Non-stationary}  &  \textbf{0.4349} &   0.5028   \\
 & \textit{Stationary} & \textbf{0.3672}  &  0.4373   \\
\textsc{twitter} & \textit{Non-stationary} & \textbf{0.4852}  &  1.2333   \\
& \textit{Stationary}  & \textbf{0.6690}  &  0.9180   \\
\bottomrule
\end{tabularx}
\end{table}

\section{Conclusion}
We proposed an evolving teleportation adaptation of the PageRank method to capture how changes in external interest influence the importance of a node.  This proposal lets us treat PageRank as a dynamical system and seamlessly incorporate changes in the teleportation vector.  
Furthermore, we demonstrated the utility of using Dynamic PageRank for predicting pageviews.  In future work, we hope to include dynamic and evolving graphs into this framework as well.  

\bibliographystyle{abbrv}
\bibliography{rossi}

\end{document}